# From Lyotropic to Thermotropic Behavior: Solvent-Free Liquid Crystalline Phases in Polymer-Surfactant-Conjugated Rod-shaped Colloidal Viruses

*Lohitha R. Hegde, Kamendra P. Sharma\**

Department of Chemistry, Indian Institute of Technology Bombay,

Mumbai-400076, India

and

*Eric Grelet\**

Univ. Bordeaux, CNRS, Centre de Recherche Paul-Pascal, UMR 5031,

F-33600 Pessac, France

\* Corresponding authors: k.sharma@chem.iitb.ac.in ; eric.grelet@crpp.cnrs.fr

**Abstract**

Filamentous bacteriophages *fd* are viral particles, highly monodisperse in size, that have been widely used as a model colloidal system for studying the self-assembly of rod-shaped particles as well as a versatile template in nanoscience. In aqueous suspensions, *fd* viruses exhibit *lyotropic* behavior, forming liquid crystalline phases as their concentration increases. Here, we report a solvent-free system displaying *thermotropic* phase behavior, achieved through covalent coupling of low molecular weight PEG-based polymer surfactant onto the *fd* virus surface. Upon lyophilization of aqueous suspensions of these polymer-grafted bacteriophages and subsequent thermal annealing, a solvent-free material is obtained, exhibiting both viscoelasticity and, notably, thermotropic liquid crystalline properties. A combination of small-angle X-ray scattering and optical microscopy experiments reveals the formation of an ordered hexagonal mesophase below 30 °C, which undergoes a melting transition into an isotropic liquid at higher temperatures. Our results demonstrate an efficient approach for converting lyotropic into thermotropic phase behavior in the columnar liquid crystalline phase of filamentous *fd* colloids. This approach paves the way for extending such functionalization to other technologically relevant rod-like systems, such as carbon nanotubes and cellulose nanocrystals, enabling the introduction of thermotropic properties in anhydrous colloidal materials.

**Introduction**

Entropically driven phase transitions of rod-like particles have been extensively investigated over the past few decades from experimental and theoretical considerations.[1-4] Onsager's seminal work has established that excluded volume interactions between rigid needle-like particles trigger the isotropic liquid-to-nematic liquid crystal phase transition at sufficiently high particle volume fractions.[5] Such rod-like particles are referred to as lyotropic colloidal systems, as their self-assembly occurs in a solvent, usually water, and is concentration-dependent. Subsequent works have revealed complex ordering beyond the nematic phase, such as smectic, columnar and other crystalline phases upon increasing rod concentration.[6-8] For lyotropic colloidal phases driven by excluded volume, the minimum of the free energy $F = U - TS$, where $U$ is the internal energy, $T$ is the temperature, and $S$ is the entropy, corresponds to a maximum of the entropy, as $F \sim -TS$. Consequently, the resultant phase behavior is inherently athermal, with the entropy $S$ being independent of temperature and solely a function of the particle concentration. Among self-organized states, the hexagonal columnar phase is particularly prevalent in dense systems, *i.e.,* at high particle packing fractions. This phase has been widely observed in various experimental systems, including imogolite,[9] carbon nanotubes,[10] nucleic acids,[11] and filamentous viruses,[8] among others. Recent numerical investigations have highlighted the challenges in stabilizing this mesophase by entropy alone,[12] suggesting the need for additional physical factors such as size polydispersity,[13] particle flexibility,[14] or surface charges.[15]

In this context, achieving a temperature-dependent, or thermotropic, hexagonal columnar organization is highly attractive. Thermally responsive materials are of major importance for numerous applications, as illustrated by molecular thermotropic liquid crystals used in cholesteric-based photonic devices[16] or discotic columnar mesophase in organic electronics.[17] A prior approach demonstrated the induction of thermotropic smectic- or lamellar-phases through the electrostatic complexation of different biomacromolecules with surfactants in anhydrous conditions.[18-20]

Here, we report an alternative solvent-free approach to generate thermotropic behavior in filamentous *fd* viruses, used as rod-like colloids. We achieve this by covalently grafting PEG-based polymer surfactants onto the virus surface, followed by a dehydration step. This solvent-free system exhibits not only viscoelastic properties, characterized by rheological measurements, but also remarkable thermotropic liquid crystalline behavior with hexagonal ordering, as confirmed by small-angle X-ray scattering (SAXS) and optical microscopy.

Our findings open a path towards tailoring the structure-property-function relationship in colloidal materials by manipulating intermolecular interactions through composition control. This approach can potentially extend to other technologically relevant rod-like objects, enabling the development of advanced thermo-responsive solvent-free liquid crystals.

## Results and Discussion

*Preparation and conjugation of NHS-activated polymer surfactant (PS-NHS) to filamentous fd bacteriophages via amine coupling*

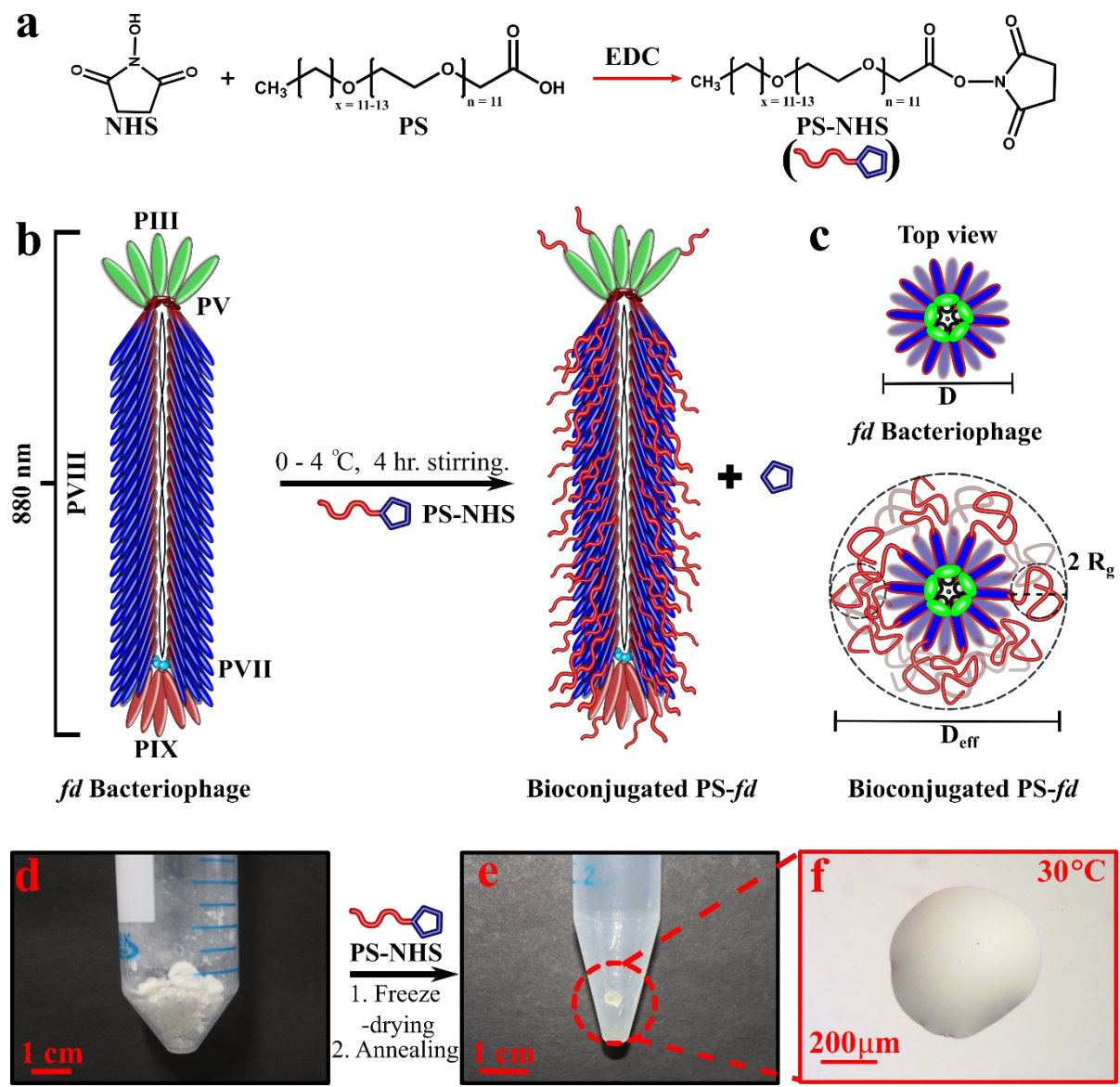

**Figure 1: Bioconjugation of *fd* bacteriophages with Polymer Surfactant (PS).** **(a)** Schematic illustration of N-hydroxysuccinimide (NHS) functionalization of the polymer surfactant (of number average molar weight $M_n \sim 690$ g. mol$^{-1}$) via esterification reaction. **(b)** Schematic representation of NHS grafting activated polymer surfactant (PS-NHS) onto *fd* bacteriophages through covalent conjugation with surface-exposed amino groups of the *fd* capsid. **(c)** Top-view comparison of wild-type *fd* bacteriophage and the bioconjugated phage (PS-*fd*), showing an increase in viral particle diameter, after polymer grafting. The effective diameter $D_{eff}$ of PS-*fd* is indicated by a dotted circle. **(d)** Photograph of lyophilized (freeze-dried) *fd* bacteriophage powder. **(e)** Photograph of a solvent-free PS-*fd* sample obtained after annealing the lyophilized PS-*fd* powder at 60 °C for 15 minutes and appearing at 25 °C as a viscous and translucent blob highlighted by a red dashed circle in an Eppendorf tube. **(f)** Bright-field optical microscopy image of a solvent-free PS-*fd* droplet deposited on a glass slide and observed at 30 °C.

N-Hydroxysuccinimide (NHS) end group functionalization of the polyethylene (PEG)-based polymer surfactant (PS, glycolic acid ethoxylate lauryl ether; $M_n \sim 690$ g mol$^{-1}$) enabled efficient grafting onto the amino groups of the surface coat proteins of filamentous *fd* bacteriophage. The NHS-functionalized polymer surfactant (PS-NHS) was synthesized via an esterification reaction using EDC (1-Ethyl-3-(3-dimethylaminopropyl) carbodiimide) as a coupling agent (**Figure 1a**, and ESI). The synthesized PS-NHS was characterized using $^1$H, $^{13}$C NMR, ATR-FTIR and UV-visible spectroscopy, which confirmed the formation of the PS-NHS ester (yield ≈ 60 %; ESI, **Figures S1** to **S4**). Our experimental system of rods is the filamentous *fd* virus, which has been widely used as a colloidal model system in soft condensed matter.[21,22] The bacteriophage *fd*, which is an almost-1-micron-length ($L = 0.88$ μm, **Figure 1**) semiflexible charged rodlike particle with a diameter of $D = 7$ nm and a persistence length of $P = 2.8$ μm, has a molecular weight of $M_w = 1.64 \times 10^7$ g. mol$^{-1}$ and is formed by single-stranded DNA around which about 2700 coat proteins displaying amino groups are helicoidally wrapped.[23,24] Bioconjugation of *fd* bacteriophages was achieved through the covalent coupling between the amino groups of the viral coat proteins and the NHS-activated PS esters, following a literature procedure with minor modifications.[25,26] Briefly, PS-NHS dissolved in 0.2 mL DMSO was added to an aqueous suspension of *fd* bacteriophages (10 mL at 1 mg.mL$^{-1}$) in phosphate buffer (PBS, 100 mM, pH 7.8), and stirred continuously for 4 hours under cold conditions (**Figure 1b**, and ESI). Considering the ~ 3000 solvent-exposed amino groups per virus particle, the polymer was added in a 10-fold molar excess relative to the accessible surface amino acids to maximize grafting density. After bioconjugation, the excess polymer was removed by extensive dialysis (MWCO 10 kDa) against phosphate buffer (PBS, 10 mM, pH 7.8) with repeated buffer change every 12 hours for 3 days. The purified bioconjugates were lyophilized (– 60 °C, 48 hours) and then annealed (60 °C, 15 min) to obtain a solvent-free liquid bioconjugate, designed as PS-*fd* (**Figure 1d-f**). The resulting bioconjugates were colorless and exhibited translucency and birefringence at room temperature, as well as viscoelastic behavior.

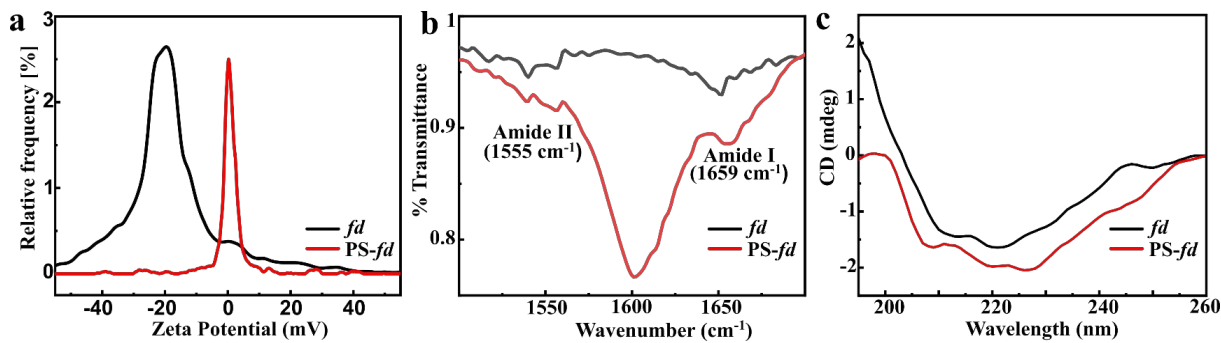

**Figure 2: Characterization of polymer surfactant-grafted *fd* bacteriophages (PS-*fd*). (a)** Zeta potential measurements of bare *fd* (black line) and bio-conjugated PS-*fd* (red line) bacteriophages show a shift in surface potential from – 22.3 mV to + 0.3 mV, consistent with polymer surfactant grafting. **(b)** Attenuated total reflectance Fourier-transform infrared (ATR-FTIR) spectra of *fd* and PS-*fd*, recorded in anhydrous conditions (25 °C), reveal characteristic amide I (1659 cm$^{-1}$) and amide II (1555 cm$^{-1}$) vibrational modes. **(c)** Circular dichroism (CD) spectra of solvent-free PS-*fd* maintain the native α-helical conformation, evidenced by characteristic negative bands at 208 nm and 228 nm.

Zeta potential measurements of *fd* bacteriophages and redispersed PS-*fd* bioconjugates in phosphate buffer (PBS, 100 mM pH 7.8) revealed a shift from – 22.3 mV to 0.3 mV, confirming successful surface modification of the viral particles (**Figure 2a**). ATR-FTIR spectroscopy of lyophilized *fd* viruses and solvent-free PS-*fd* liquids displayed characteristic absorption peaks at 1659 cm$^{-1}$ (amide I) and at 1555 cm$^{-1}$ (amide II) bands, consistent with the preservation of α-helical secondary structures even after bioconjugation (**Figure 2b**). Note that the strong minimum observed at 1600 cm$^{−1}$ for PS-*fd* can be attributed to the convolution of multiple vibrational sources in the bioconjugated PS-*fd*. These include the *β*-sheet amide I stretches of the capsid protein, carbonyl vibrational modes from the tightly packed viral DNA bases, and the newly introduced amide/carboxylate C=O groups originating from polymer-surfactant (see ESI, **Figure S3**) conjugation of *fd*, resulting in composite spectral features in this region rather than a single structural signature. These observations were consistent with the literature studies.[27] Circular dichroism spectra of PS-*fd* showed characteristic peaks for the α-helix of the main coat protein similar to native *fd* bacteriophages, further verifying structural preservation of the protein secondary structure, as supported by ATR-FTIR data (**Figure 2c**).

*Grafting density and conformation of polymer surfactants grafted on fd bacteriophages*

The grafting density and conformation of polymer surfactants attached to *fd* bacteriophages were characterized. Generally, grafting-*to* methods produce lower polymer grafting densities compared to grafting-*from* approaches, for which polymerization is initiated *in situ* from the surface.[28] To quantify the grafting density (defined as the number of polymer surfactant chains per unit area conjugated to each *fd* bacteriophage), we performed differential refractive index (*dn/dc*) measurements (see ESI and **Figure S5**). Since the differential refractive index *dn/dc* is directly proportional to mass density, its value varies between native and bio-conjugated systems.[25] Distinct *dn/dc* values were obtained for the different systems: 0.194 for native *fd*, 0.201 for PS-*fd* conjugates, and 0.143 for PS-NHS. These measurements revealed an experimental grafting density of approximately $N_{exp}$ = 1003 polymer surfactant chains per *fd* bacteriophage (corresponding to 0.055 PS molecules per nm$^2$). Upon conjugation, the grafted polymer chains adopt distinct conformations depending on their surface density: coil-like or mushroom conformation at low grafting density and brush-like regime with stretched chains at high grafting density. Considering the grafting-*to* approach and therefore assuming a polymer coil conformation, the theoretical maximum grafting capacity onto the surface of the bacteriophage can be calculated using the following relation based on simple geometrical arguments:[26]

$$N_{cp} = \frac{\pi(D+2R_g)L}{\pi R_g^2} \quad (1)$$

where $N_{cp}$ is the highest amount, corresponding to close-packed particles of chains grafted per virus, $L$ and $D$ are the virus length and diameter, respectively, and $R_g$ is the polymer surfactant radius of gyration, determined for an ideal linear chain as:

$$R_g = \sqrt{\left(\frac{C_\infty n_b}{6}\right)} \times l \qquad (2)$$

With $C_\infty \cong 7$, the characteristic ratio, $n_b = 50$, is the number of bonds in the polymer backbone (**Figure 1a**), and $l = 0.15$ nm is the C-C single bond length.[29] For our polymer surfactant PS, this leads to $R_g \approx 1.2$ nm. Substituting this value in equation (1), the maximum theoretical grafting density was determined to be $N_{cp} = 5960$ chains per virus, *i.e.*, $N_{exp} < N_{cp}$. The experimental value ($N_{exp} = 1003$) being significantly lower than this theoretical maximum confirms an intermediate grafting density regime, consistent with the expected mushroom-like polymer conformation. The successful grafting was further verified by thermal analysis, which showed an increase in the melting transition temperature upon PS conjugation (**Figure 3**). This observation contrasts sharply with the behavior expected for mixtures of ungrafted PS with *fd* inclusions, where thermodynamic principles predict a decrease in the melting transition temperature, as expected for binary mixtures.

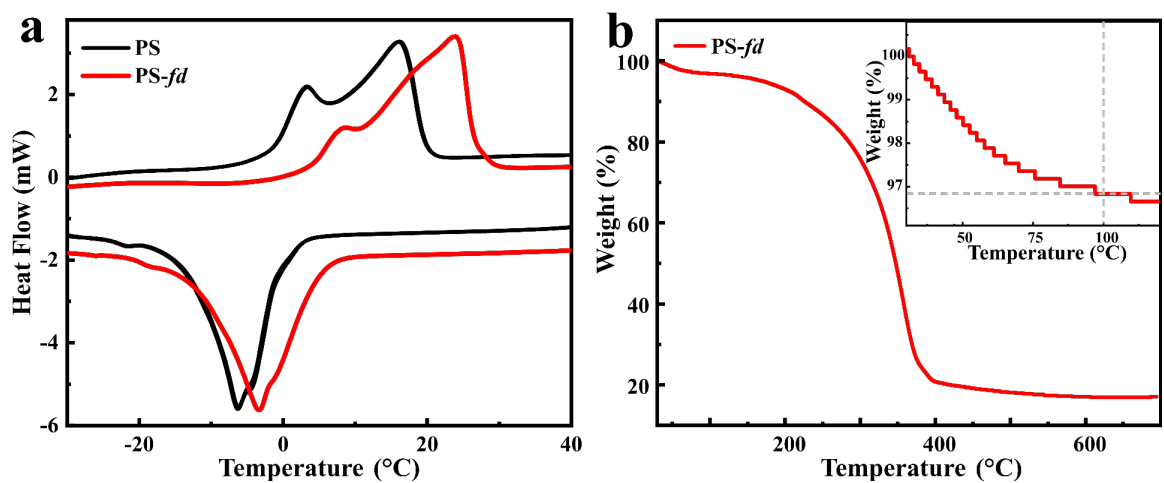

**Figure 3: Thermal characterization of polymer surfactant and bacteriophage bioconjugates.** **(a)** Differential scanning calorimetry (DSC) of the polymer surfactant (PS), and solvent-free bacteriophage conjugates (PS-*fd*). The heating and cooling second cycles

indicate that PS exhibits a crystallization temperature of – 6.2 °C, and two melting transitions at 3.3 °C and 16.4 °C. In contrast, PS-*fd* displays an exothermic crystallization at – 3.3 °C and a broad endothermic transition in the range of 25-30 °C. **(b)** Thermogravimetric analysis (TGA) of PS-*fd* reveals an initial mass loss of 3% (see inset) below 150 °C, attributed to residual water content in the sample. Given the negligible amount of residual solvent, the PS-*fd* system can be considered effectively as solvent-free. The primary thermal decomposition occurs near 350 °C.

*Phase behavior and microstructural studies of the solvent-free bioconjugated PS-fd*

To investigate thermal phase behavior, differential scanning calorimetry (DSC) was performed on both PS and solvent-free PS-*fd* samples (**Figure 3a**). The PS thermogram reveals a crystallization temperature of – 6.2 °C and two melting transitions at 3.3 °C and 16.4 °C. These multiple melting points are likely attributable to the molecular weight polydispersity of the PS, consistent with the previous reports.[30] While PS-*fd* thermograms exhibit qualitative similarities to that of PS, all transition temperatures are shifted to higher values - a direct consequence of bioconjugation to *fd* surfaces. PS-*fd* displays a broad exothermic crystallization peak at – 3.3 °C, likely arising from PEG-PEG and alkyl-alkyl interactions within the PS chains. A subsequent broad endothermic melting transition, $T_m$, occurs in the range of 25–30 °C, and may correspond to an ordered phase to isotropic phase transition specific to the solvent-free PS-*fd* system. Thus, while the thermal transitions in PS-*fd* are primarily governed by the PS component, the changes induced by bioconjugation significantly influence its thermal behavior. Moreover, the similarity between the melting transitions of PS-*fd* and the pure PS suggests that long-range intramolecular interactions between polymer chains grafted onto the bacteriophage surface contribute to the observed viscous and translucent mesomorphic state (**Figure 1e**).

Thermogravimetric analysis (TGA) of the PS-*fd* bioconjugates reveals a residual water content of about 3 wt. %, corresponding to roughly to $5.59 \times 10^5$ water molecules per *fd*-bacteriophage. This water content is smaller by more than a factor of 2 than the 7 wt. % typically required to support proper protein mobility and function.[31,32] Due to the negligible residual solvent content, the PS-*fd* can be effectively considered as a solvent-free system. Furthermore, the solvent-free PS-*fd* system exhibits minimal weight loss near 220 °C and a major decomposition at ≈ 350 °C (**Figure 3b**).

The viscoelastic properties of the PS-*fd* system were investigated using oscillatory shear rheology. Temperature-dependent viscosity measurements conducted between 18 °C and 80 °C, at a constant shear rate of 15 s$^{-1}$, revealed a gradual decrease in viscosity from 49.4 Pa.s to 4.9 Pa.s, with increasing temperature (**Figure 4a**).

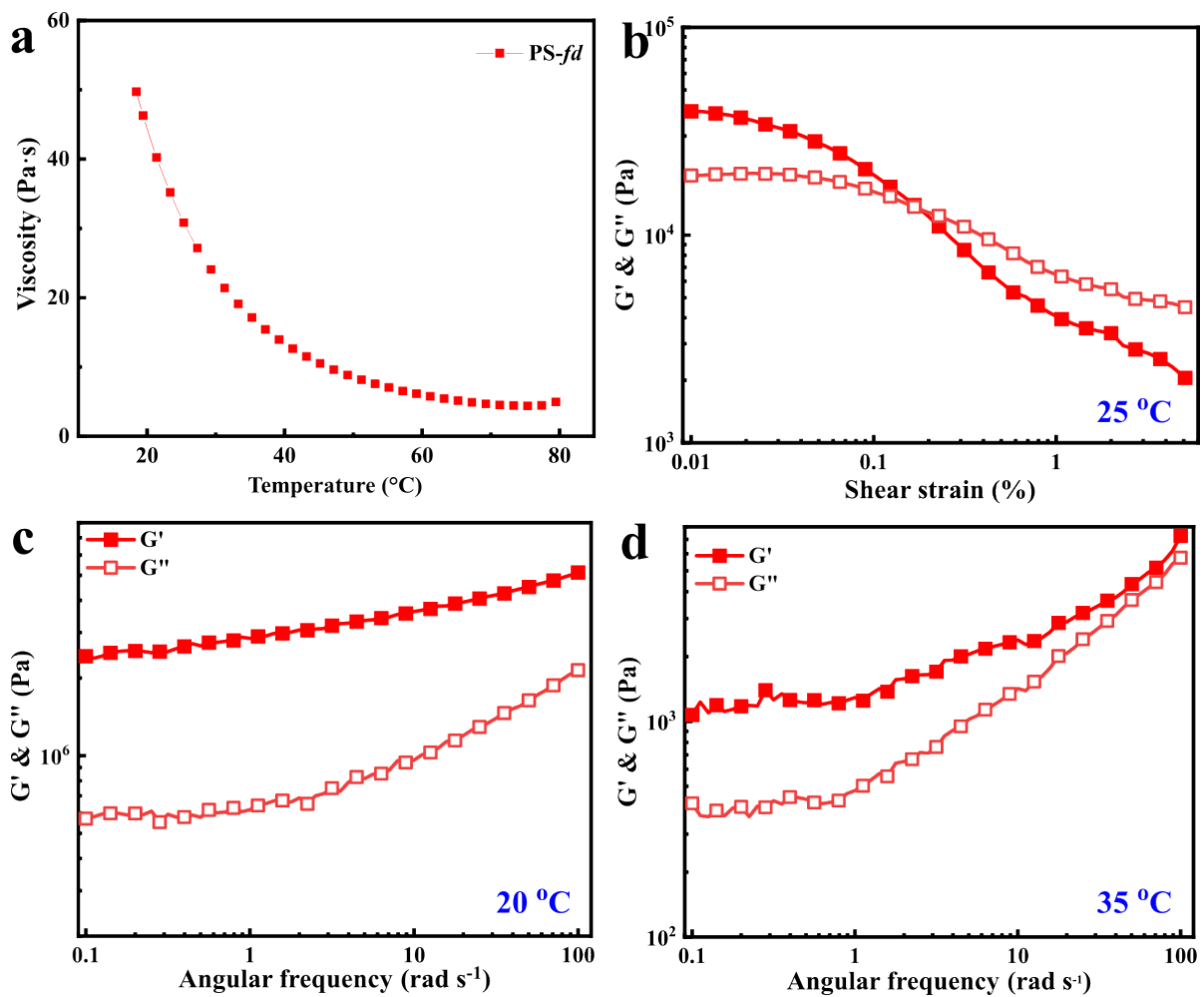

**Figure 4: Rheology of the solvent-free PS-*fd* system. (a)** Temperature-dependent viscosity measurements performed in the temperature range of 18–80 °C at a constant shear rate of 15 s$^{-1}$. **(b)** Amplitude sweep experiment performed at 25 °C and at an angular frequency of ω =10 rad.s$^{-1}$, within shear strain ranging from 0.01 to 5 %. In the low-strain regime (0.01–0.1%), the storage modulus G′ (full symbols) exceeds the loss modulus G″ (open symbols), indicating an elastic solid-like behavior of the PS-*fd* system. **(c,d)** Frequency sweep experiment performed

within the linear viscoelastic regime (strain of 0.01 %), over the range from 0.1 to 100 rad.s$^{-1}$, at **(c)** 20 °C and **(d)** 35 °C. In both cases, the storage modulus G′ remains higher than the loss modulus G″, confirming an elastic solid-like feature. An increase in temperature leads to a reduction of about three orders of magnitude in both G′ and G″.

Amplitude/strain sweep experiments performed at 25 °C revealed a linear viscoelastic region in the range of 0.01 to about 0.1% shear strain. Within this range, the storage modulus (G′) exceeds the loss modulus G″, indicating an elastic solid-like behavior of the solvent-free PS-*fd* (**Figure 4b**). Subsequently, frequency sweep experiments were carried out at a constant strain of 0.01 % over the range of 0.1 to 100 rad.s$^{-1}$, at 20 °C and 35 °C, *i.e.,* below and above the melting transition temperature $T_m$, respectively. At 20 °C, G′ and G″ are approximately 2.43 × 10$^6$ Pa and 5.7 × 10$^5$ Pa, respectively, indicating a prevailing viscoelastic solid-like behavior (G′ > G″), consistent with the presence of structural organization (**Figure 4c**). In contrast, at 35 °C, both moduli decrease by about three orders of magnitude, with G′ ≈ 1.1 × 10$^3$ Pa and G″ ≈ 0.4 × 10$^2$ Pa (**Figure 4d**). Consequently, the loss tangent (tan δ = G″/ G′) increases from 20 °C to 35 °C, reflecting a transition from solid-like to more fluid-like behavior.[33,34] This change is attributed to a phase transition in the PS-*fd* system occurring at $T_m$ around 30 °C, consistent with calorimetry (DSC) results.

Small-angle X-ray scattering (SAXS) was performed at various temperatures to investigate the microstructural phase behavior of the solvent-free PS-*fd* system and compare it with that of the polymer surfactant, PS (**Fig. 5** and **Fig. S6**). SAXS measurements were carried out during both cooling and heating cycles to examine hysteresis effects as well as the structural transitions from pre-annealed samples (70 °C). In both heating and cooling cycles, the PS-*fd* bioconjugate exhibits, above 35 °C, a monotonically decaying scattering profile with a single broad peak at a scattering vector $q_3 ≈ 0.14$ Å$^{-1}$, corresponding to a real-space distance $d_3 ≈ 2\pi/q_3$ ≈ 4.5 nm, characteristic of an isotropic liquid phase (**Figures 5b, c** and **d**). A similar broad peak is also observed for pure PS at $q ≈ 0.15$ Å$^{-1}$ above its melting transition temperature (**Figure**

**S6**). Given the high PS content in the PS-*fd* sample, the $q_3 \approx 0.14$ Å$^{-1}$ peak can be attributed to scattering from the PS corona surrounding the *fd* bacteriophage. Upon cooling below the melting temperature $T_m \approx 30$ °C, the $q_3$ peak sharpens (**Figure 5f**), indicating increased ordering of the PS component. Simultaneously, two distinct Bragg reflections appear at positions $q_1$ and $q_2$, arising from the long-range positional ordering of the *fd* viruses and the alignment of the PS chains perpendicular to the *fd* long axes (**Figure 5**). This indicates cooperative organization of both the polymer-surfactant and the bioconjugated virus components upon transition to the ordered phase.

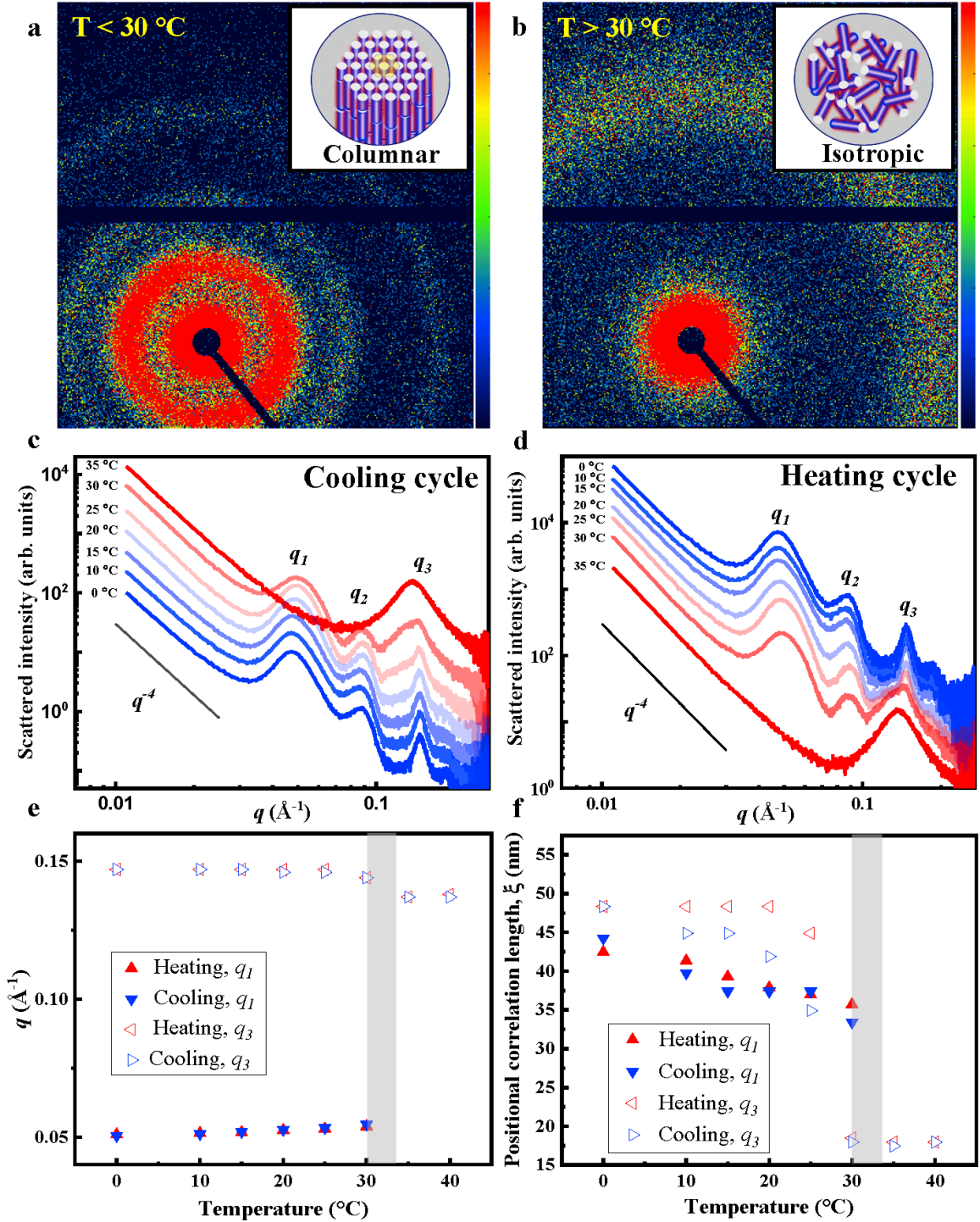

**Figure 5:** Temperature-dependent SAXS measurements on the solvent-free PS-*fd* bioconjugate. **(a)** Two-dimensional SAXS patterns recorded below the melting temperature ($T_m \approx 30$ °C) showing hexagonal columnar ordering. **(b)** The corresponding pattern above $T_m$ of PS-*fd* is in an isotropic liquid-like phase. Schematic illustrations of each structure are shown as insets. 1D scattered profiles of PS-*fd* samples recording while **(c)** cooling and **(d)** heating the sample between 0 °C and 35 °C. For $T > T_m$, a single broad peak at position $q_3$ indicates the absence of long-range positional order. Below $T_m$, two Bragg reflections at $q_1$ and $q_2$ appear with the characteristic ratio $q_2 / q_1 = \sqrt{3}$ as a hallmark of hexagonal symmetry perpendicular

to the long axis of the viral rods. Temperature evolution of the **(e)** position of the first $q_1$ and third $q_3$ peaks and **(f)** the corresponding positional correlation length, $\xi$, obtained from the full width at half-maximum (FWHM) of the Bragg reflection via $\xi=2\pi/$FWHM.

In detail, the position of the second peak $q_2$ relative to the position of the first Bragg peak at $q_1 \sim 0.05$ Å$^{-1}$ follows a ratio of 1:√3, consistent with (10) and (11) Miller indices. These peaks are characteristic of hexagonal positional order in the plane normal to the long axis of the viral rods. It is important to note that pure PS tends to form a lamellar phase below its melting temperature, around 16 °C, as evidenced by the SAXS data in **Figure S6**. The formation of the columnar hexagonal phase in the PS-*fd* system at significantly higher temperatures is therefore attributed to the synergistic coupling between the anisotropic shape of the filamentous bacteriophage and the structuration of the surrounding polymer surfactant grafted on the virus surface. In solvent-free PS-*fd*, the first-order Bragg peak at $q_1$ corresponds to a lattice spacing $d_1 \approx 14.5$ nm, calculated for a hexagonal lattice using the relation $d_1 = \frac{4\pi}{\sqrt{3}\,q_1}$. This spacing reflects the average center-to-center distance between neighboring *fd* bacteriophages arranged in a hexagonal array. This value can be directly compared with the effective diameter of the PS-*fd* particle, which can be calculated as the bare diameter of the *fd* virus, $D$, plus twice the diameter (estimated to be $2.R_g$) of the PEG-based polymer surfactant:

$$D_{\text{eff}} = D + 4\,R_g \tag{3}$$

Taking $R_g \approx 1.2$ nm as determined above, this results in $D_{\text{eff}} \approx 11.8$ nm, which is consistent with the observed hexagonal lattice parameter $d_1$ of the ordered microstructure in solvent-free PS-*fd* below 30 °C considering the relative uncertainty in determining the PS radius of gyration. This supports the interpretation that PS-*fd* forms a highly packed structure in the ordered phase, consistent with the expectations for a nearly solvent-free system driven by predominant steric repulsion between the PS shells of adjacent virus particles. It is also worth noting that the structure factor associated with the organization of micrometer-long rod-like viruses in the isotropic phase above the melting temperature $T_m$ likely appears at much lower $q$-values than the $q$-range probed here. Interestingly, in the low-$q$ regime, a $q^{-4}$ scattering behavior – characteristic of the Porod regime – is clearly distinguished (**Figures 5c, d**), indicating the presence of sharp, well-defined interfaces in the system. These sharp interfaces may arise at the particle level, *i.e.*, between the *fd* virus and the polymer surfactant, and at larger structural scales (and therefore lower $q$) in the grain boundaries separating PS-*fd* microdomains, where the residual water content may be localized.

The thermal transitions observed via DSC and SAXS for solvent-free PS-*fd* samples revealed hexagonal-to-isotropic liquid transitions above the melting temperature of 30 °C. Temperature-dependent optical microscopy (**Figure S7**) was performed to corroborate the SAXS and DSC results and to investigate the nucleation and growth of the ordered microstructure. Samples were annealed for 10 min at different temperatures on a heating stage prior to imaging. Initially, PS-*fd* samples were annealed at 40 °C, cooled gradually to 0 °C, and then heated to 40 °C at a rate of 10 °C.min$^{-1}$. During the cooling cycle, the samples exhibited an isotropic liquid-like state at 30 °C, with no birefringence observed due to the melting of the PS-grafted *fd*. Below the melting transition (≈ 28 °C), nucleation of highly ordered birefringent textures occurred, followed by further growth upon continued cooling **(Figure S7)**. In the heating cycle, birefringence gradually diminished and disappeared around 32 °C, *i.e.*, 4 °C higher than the transition during cooling (**Figure S7**). This behavior aligns with our SAXS results, where the hexagonal-to-isotropic transition occurred at higher temperatures for the heating cycle. In the case of pure PS, spherulites with birefringent textures nucleated near 15 °C, and disappeared above the melting transition (≈ 20 °C, **Figure S8**). The thermal phase behavior of PS-*fd*, compared to pure PS, indicates that the phase transitions in PS-*fd* arise primarily from the surface grafting of polymer surfactant chains onto the bacteriophages. The observed optical microscopy-based thermal hysteresis is observed, as evidenced by the temperature shift between heating and cooling cycles, is consistent with the phase transitions of polymeric systems that transform from elastic solids to viscous liquids and *vice versa*. This can be related to the nucleation and growth of PS chains into crystalline structures, as previously reported in the literature.[30]

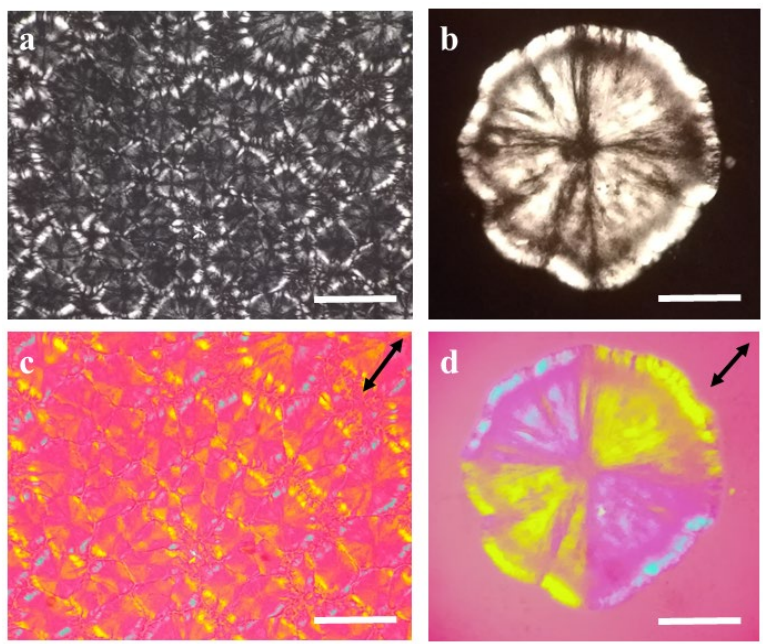

**Figure 6: Polarizing optical microscopy texture** (top left) and single domain (top right) of a solvent-free PS-*fd* droplet sandwiched between a glass slide and cover-slip and observed below $T_m$ **(a,b)** and with the addition of a full-wave retardation plate **(c,d)** with its slow axis indicated by a black double arrow. The retarded optical path results in fast (first-order yellow) and slow (second-order blue/violet) axes in an isotropic medium (magenta background), indicating a columnar liquid crystalline phase configuration.[35] The scale bars are 100 μm **(a,c)** and 20 μm **(b,d)**, respectively.

As shown in **Figure 6**, the PS-*fd* samples display pronounced birefringence under optical microscopy, exhibiting fan-shaped or pleated textures characteristic of hexagonally ordered structures, consistent with previous reports.[8,11] In order to determine the orientation of the columns within these fan-shaped domains as shown in **Figure 6b**, a full-wave retardation plate was used to identify the slow and fast vibration axes of this uniaxial birefringent material.[35] When the slow vibration axis of the sample aligns with the slow axis of the λ-plate, the additive retardation effects result in higher-order interference colors (blue/violet color).

Conversely, when the sample's fast axis is parallel to the slow axis of the retardation plate, it results in lower-order interference colors (yellow color). Given the positive birefringence of *fd* virus system (*i.e.*, $n_e>n_o$, implying $v_e<v_o$, where v denotes the speed of light in the anisotropic medium), the slow axis corresponds to the extraordinary axis ($n_e$) and the fast vibration axis to the ordinary index ($n_o$). Thus, in the PS-*fd* system, the fan-shaped texture observed in **Figure 6d** arises from concentrically wrapped virus particles, self-assembled into columns, and widely observed for hexagonal mesophases.

The phase behavior of *fd* bacteriophages in a solvent-free polymeric environment warrants further discussion. In solvent-free liquids, intermolecular interactions are governed primarily by steric repulsions and van der Waals forces.[36] While steric repulsion arises from excluded volume effects, van der Waals interactions depend on refractive index differences. In this solvent-free system comprising monodisperse *fd* bacteriophages (~ 18 % v/v as calculated using SAXS) covalently modified with ≈ 0.7k $M_n$ PEG-based polymer surfactants at low-to-medium grafting densities, a stable hexagonal mesophase emerges. Notably, such an ordered structure has not been reported for other recently developed solvent-free systems.[18] Moreover, the polymer-surfactant alone self-organizes into a lamellar phase at significantly lower temperatures. Therefore, it can be argued that the hexagonal ordering in this solvent-free and highly crowded environment stems from the combination of three factors: (1) the rod-like anisotropy of *fd* bacteriophages, (2) the conformational flexibility of short PEG polymer-surfactant chains, and (3) a balance between steric repulsions and weak van der Waals interactions between the PS-*fd* bioconjugated particles.

**Conclusions**

In summary, we have developed an NHS coupling-based strategy for covalently grafting low molecular weight polyethylene glycol (PEG)-based polymer surfactants onto the surface of filamentous *fd* bacteriophages in aqueous solution. Extensive lyophilization of the resulting bioconjugates, followed by thermal annealing, yields a solvent-free viscoelastic material stable at room temperature, as determined by calorimetric and rheological analyses. Small-angle X-ray scattering and polarized optical microscopy reveal the formation of an ordered hexagonal mesophase below ≈ 30 °C, which upon heating melts into an isotropic liquid phase. Our study thus provides important insights into the phase behavior and structure-property relationships of densely packed rod-like particles. Notably, this versatile methodology can be extended to other technologically relevant anisotropic colloids, such as carbon nanotubes and cellulose nanocrystals, to engineer new classes of colloidal liquid crystals with *thermo*responsive phase behavior and tunable material properties.

**Electronic Supplementary Information** (ESI) available.

**Author contributions**

Lohitha R. Hegde: investigation, formal analysis, methodology, visualization, writing – original draft, writing – review and editing.

Kamendra P. Sharma: conceptualization, formal analysis, funding acquisition, methodology, visualization, writing – original draft, writing – review and editing.

Eric Grelet: conceptualization, formal analysis, investigation, methodology, visualization, writing – original draft, writing – review and editing.

**Data availability**

All the data supporting the findings of this study are included in the article and its Electronic Supplementary Information file and will be made available on request.

**Conflicts of interest**

There are no conflicts to declare.

**Acknowledgements**

L.R.H. and K.P.S. acknowledge support for SAXS studies to S.A.I.F., IIT Bombay, TEM imaging to Raju K. Singh, and Central facility, Chemistry department, IIT Bombay. K.P.S. acknowledges financial support from DST, grant number RD/0120-SERB000-035, and Seed funding for Collaboration and Partnership Projects (SCPP) grant RD/0523-IOE00I0-075 IoE funding from IIT Bombay.